\documentclass[showpacs,preprintnumbers]{revtex4}

\usepackage{graphicx}
\usepackage{dcolumn}
\usepackage{bm}

\begin{document}

\title{Six-fold in-plane magnetic anisotropy in Co-implanted ZnO (0001)}

\author{Numan Akdo\v{g}an}
\altaffiliation{Author to whom correspondence should be addressed.
E-mail address: akdogan@gyte.edu.tr}
\author{Bulat Rameev}
\altaffiliation{Also at Kazan Physical-Technical Institute, 420029
Kazan, Russia}
\author{S\"{u}meyra G\"{u}ler}
\author{Osman \"{O}zt\"{u}rk}
\author{Bekir Akta\d{s}}
\affiliation{Gebze Institute of Technology, Department of Physics,
41400 Kocaeli, Turkey}

\author{Hartmut Zabel}
\affiliation{Institut f\"{u}r
Experimentalphysik/Festk\"{o}rperphysik, Ruhr-Universit\"{a}t
Bochum, D-44780 Bochum, Germany}

\author{Rustam Khaibullin}
\author{Lenar Tagirov}
 \altaffiliation{Permanent address: Kazan State University, 420008 Kazan, Russia}
\affiliation{Kazan Physical-Technical Institute of RAS, 420029
Kazan, Russia}

\date{\today}

\begin{abstract}
Magnetic anisotropies of Co-implanted ZnO (0001) films grown on
single-crystalline Al$_2$O$_3$ ($11\overline{2}0$) substrates have
been studied by ferromagnetic resonance (FMR) technique for
different cobalt implantation doses. The FMR data show that the easy
and hard axes have a periodicity of $60^\circ$ in the film plane, in
agreement with the hexagonal structure of the ZnO films. This
six-fold in-plane magnetic anisotropy, which is observed for the
first time in ZnO-based diluted magnetic semiconductors, is
attributed to the substitution of cobalt on Zn sites in the ZnO structure,
and a clear indication for long range ferromagnetic
ordering between substitutional cobalt ions in the
single-crystalline ZnO films.
\end{abstract}


\maketitle

The transition metal (TM)-doped ZnO has attracted a significant
amount of interest as a potential oxide-based diluted magnetic
semiconductor (DMS) material for implementation in novel spintronic
devices. After theoretical predictions of room temperature
ferromagnetism in TM-doped ZnO
\cite{DietlSci00,SatoJJAP00,SatoJJAP01}, a number of experimental
works on these systems has been done
\cite{UedaAPL01,PeartonMSE03,PrellierJPCM03,TuanPRB04,PeartonSST04,PeartonJVST04,
LiuJMS05,JanischJPCM05,LiuJAP06,PacuskiPRB06,ZhangJPCM07,BehanPRL08,AkdoganJPDAP08,AkdoganJAP09}.
Actually, some of these studies indeed claim ferromagnetic signals
above room temperature. However, the main unresolved question is
whether the observed ferromagnetism originates from uniformly
distributed TM elements in the ZnO host matrix or whether it is due
to the precipitation of metallic ferromagnetic clusters. Recently,
we have reported that the implantation of cobalt ions into the
nonmagnetic ZnO film causes intrinsic ferromagnetism at room
temperature and simultaneously creates \emph{n}-type charge carriers
without additional doping \cite{AkdoganJPDAP08,AkdoganJAP09}. We
have observed magnetic dichroism at the Co $\emph{L}_{2,3}$ and O
\emph{K} edges at room temperature as well as the multiplet
structure in x-ray absorption spectra around the Co $\emph{L}_3$
edge, that supports the intrinsic nature of the observed
ferromagnetism in Co-implanted ZnO films. Furthermore, we have found
that the magnetic moment per substituted cobalt in ZnO is about 2.81
$\mu_B$ which is very close to the theoretical expected value of 3
$\mu_B/Co$ for Co$^{2+}$ in its high spin state.

In this letter, we report on the six-fold in-plane magnetic
anisotropy in the Co-implanted ZnO films, observed for the first
time by room-temperature FMR technique. The ZnO (0001) thin films
were grown on sapphire ($11\overline{2}0$) substrates by RF (13.56
MHz) sputtering of a ZnO target \cite{AyASS03}. After post-growth
annealing, the ZnO samples were implanted with 40 keV Co$^+$ ions to
the implantation doses in the range of $0.25-2.00\times10^{17}
ions\cdot cm^{-2}$. More details on growth, structural, electronic
and magnetic properties were published elsewhere
\cite{AkdoganJPDAP08,AkdoganJAP09}.

\begin{figure}[!h]
\includegraphics[width=0.5\textwidth]{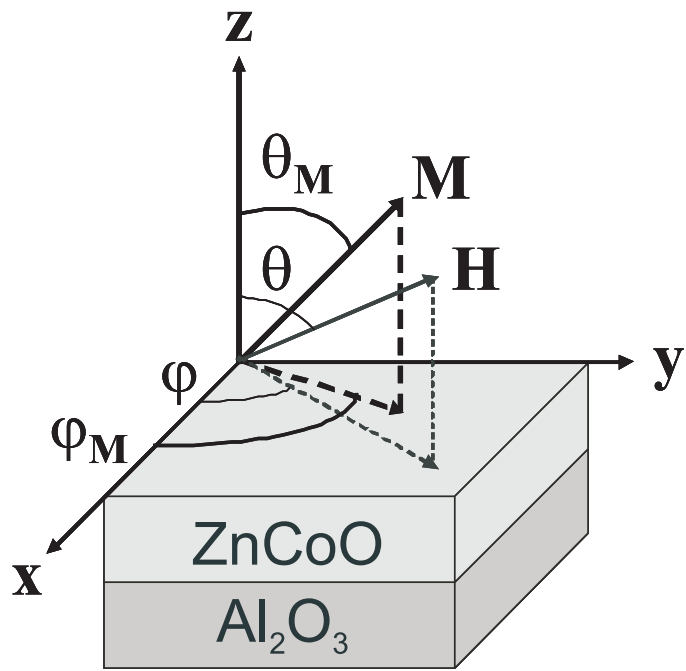}
\caption{\label{1-geometry} The coordinate system for FMR
measurements of Co-implanted ZnO films.}
\end{figure}

FMR measurements were carried out using a commercial Bruker EMX
electron spin resonance (ESR) spectrometer operating in X-Band (9.8
GHz) at room temperature. Angular dependencies of FMR spectra have
been recorded with the static magnetic field rotated either in the
plane of the samples (in-plane geometry - $\theta$=90$^\circ$,
$\varphi$-varied) or rotated from the sample plane to the normal
(out-of-plane geometry - $\theta$-varied, $\varphi$-fixed). The used
coordinate axes and relative orientation of applied external
magnetic field (\emph{H}) and magnetization vector (\emph{M}) are
illustrated in Fig.~\ref{1-geometry}.

\begin{figure}[!h]
\includegraphics[width=1\textwidth]{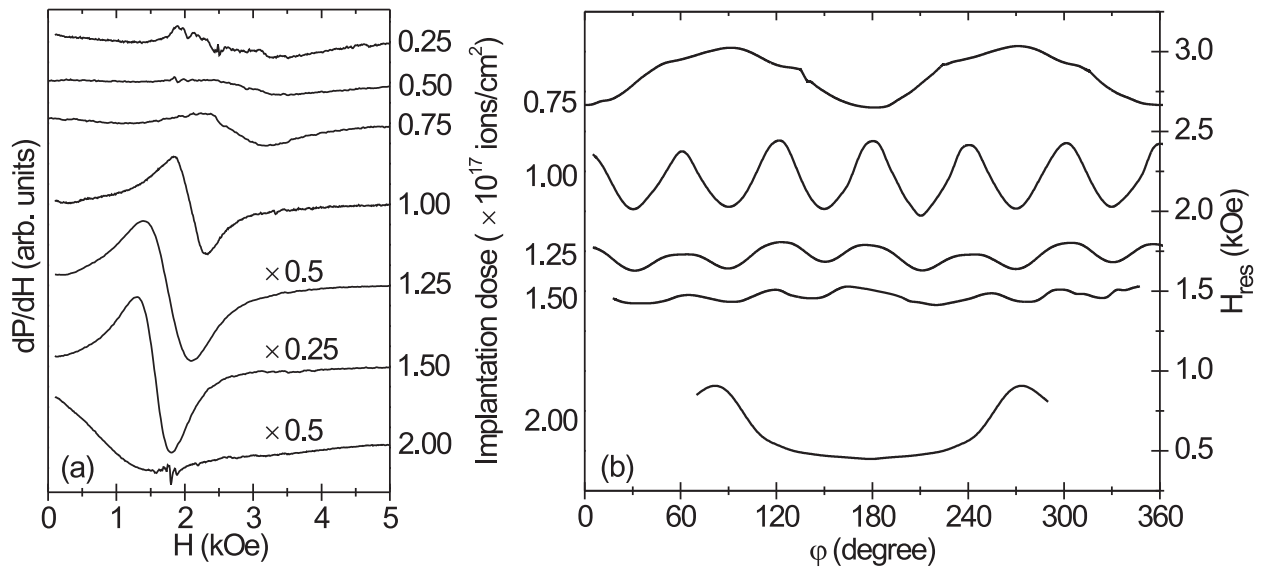}
\caption{\label{2-inplane} FMR spectra at the in-plane geometry (a)
and the angular dependence of the resonance fields (b) for different
cobalt implantation doses.}
\end{figure}

In Fig.~\ref{2-inplane}, we present in-plane FMR spectra of
Co-implanted ZnO films and angular dependence of the in-plane
resonance fields at room temperature for different implantation
doses. The resonance field exhibits oscillatory behavior as a
function of the azimuthal angle. The maximum and minimum values of
the resonance fields correspond to the hard and easy directions for
the magnetization, respectively. The periodicity of the easy and
hard axes depends on the implantation dose. As seen in
Fig.~\ref{2-inplane}, at implantation doses lower than
$0.75\times10^{17} ions\cdot cm^{-2}$, the FMR signal is very weak.
For the dose of $0.75\times10^{17} ions\cdot cm^{-2}$, the FMR
signal is pronounced and a two-fold in-plane magnetic anisotropy is
observed with a very small contribution from a six-fold anisotropy.
The two-fold in-plane magnetic anisotropy is related to cobalt
nanoparticles forming a cobalt rich layer in the sapphire substrate,
close to the ZnO/Al$_2$O$_3$ interface \cite{AkdoganJPDAP08}.
Indeed, when the crystalline Al$_2$O$_3$ is implanted with cobalt
ions, Co nanoparticles with the hexagonal structure are aligned with
their c-axis parallel to the c-axis of the host sapphire
\cite{MeldrumNIMPRB03}. Thus, the Al$_2$O$_3$ matrix provides a
magnetic anisotropy to cobalt nanoclusters \cite{LeeAPL06}. In our
case, the c-axis of the host Al$_2$O$_3$ is in the sample plane.
Therefore, we infer that the two-fold in-plane magnetic anisotropy
results from cobalt nanoparticles in agreement with previous studies
\cite{MeldrumNIMPRB03,LeeAPL06}. For the dose range of
$1.00-1.50\times10^{17} ions\cdot cm^{-2}$, the corresponding FMR
data show that the easy and hard axes have a periodicity of
$60^\circ$ in the film plane, in agreement with the hexagonal
structure of the ZnO films. This six-fold in-plane magnetic
anisotropy is attributed to the substitution of cobalt on Zn sites
in the ZnO layer and it is a clear indication for long range
ferromagnetic ordering between substitutional magnetic cobalt ions
in the ZnO crystal structure. At the highest implantation dose of
$2.00\times10^{17} ions\cdot cm^{-2}$, a two-fold in-plane magnetic
anisotropy appears again. This means that for the highest dose, not
only substituted cobalt ions but also metallic cobalt clusters are
present in the ZnO layer, in accordance with the results published
in Ref.\cite{AkdoganJAP09}. Thus, for the highest implantation dose
the resonance signal is ascribed to an overall response of the metal
cobalt nanoparticles in both the ZnO layer and the sapphire
substrate. In this respect, a gradual decrease of the magnitude of
six-fold anisotropy for the dose of $1.5\times10^{17} ions/cm^{2}$)
is noteworthy. Therefore, one can expect the formation of very small
cobalt clusters in ZnO layer below the dose of $2.00\times10^{17}
ions/cm^{2}$. Thus, the maximum amplitude of the six-fold
anisotropy, revealed at the dose of $1.00\times10^{17} ions/cm^{2}$,
reflects the limit where the highest concentration for the
substitutional cobalt phase in ZnO is reached. For higher doses,
formation of the extrinsic ferromagnetic phase due to Co clusters
starts.

\begin{figure}[!h]
\includegraphics[width=1\textwidth]{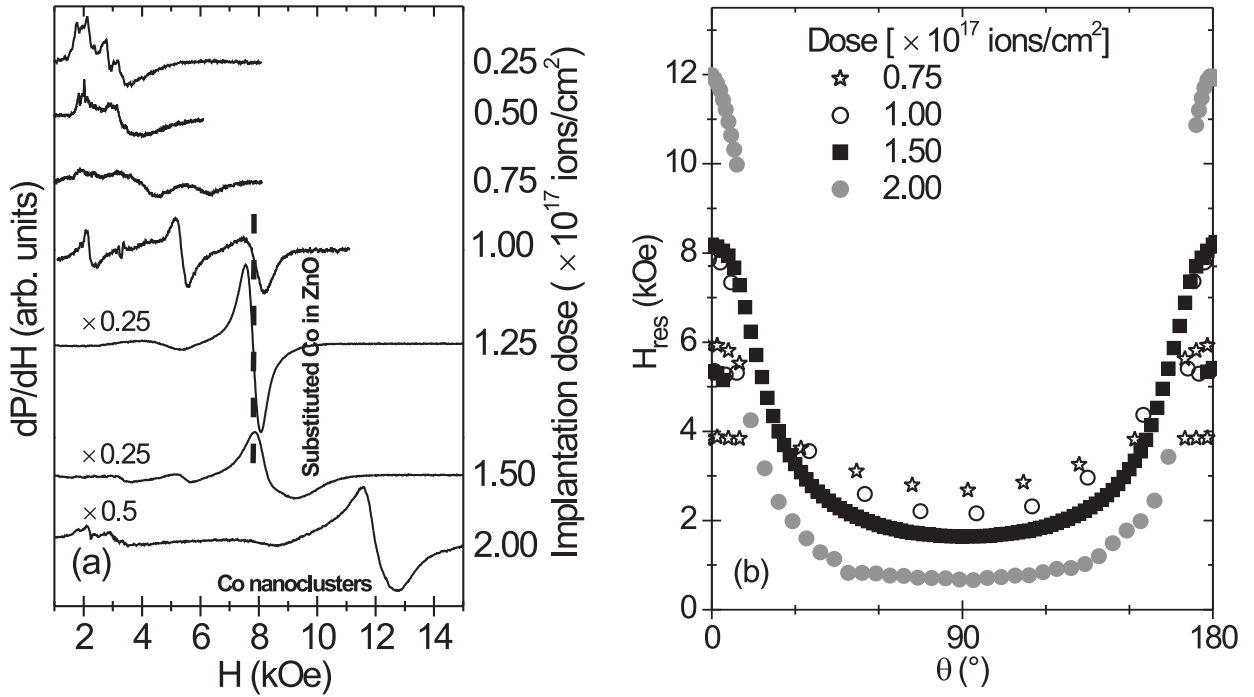}
\caption{\label{3-outplane} Out-of-plane FMR spectra (a), and the
resonance field of FMR signal as a function of polar angle (b), for
different cobalt implantation doses}
\end{figure}

Out-of-plane FMR spectra and the FMR resonance fields as a function
of polar angle ($\theta$) are shown in Fig.~\ref{3-outplane} for
different cobalt implantation doses. At low doses the resonance
signal is very weak. Increase of the implantation dose results in
appearance of the FMR signal, characterized by the six-fold in-plane
magnetic anisotropy and attributed to the substitutional cobalt
phase in the ZnO layer. In the out-of-plane geometry, this signal
consists of a single line for the parallel orientation and two lines
for the perpendicular orientation of the ZnO film with respect to
the dc magnetic field. The splitting of this signal into two modes
is related to the non-homogeneous profile of the cobalt
concentration across the film thickness. It is well known that in
systems where a gradient of the magnetization as well as a
difference in the surface/interface anisotropies exists, non-uniform
FMR modes could be observed \cite{PortisAPL63,WigenTSF84}.
Therefore, the splitting into two modes, observed in the
perpendicular orientation, is explained by excitation of the
surface/interface located modes (at lower field/higher fields,
respectively). It should be noted that in the dose range of
$1.00-1.50\times10^{17} ions/cm^{2}$, where the six-fold anisotropy
dominates, the resonance fields and intensity of the low-field mode
change only slightly, while the intensity of the high-field mode
gradually increases. The latter reflects increased influence of the
cobalt metal nanoparticles on the overall FMR signal upon increasing
the implantation dose. A strong FMR signal with high effective
magnetization, observed at the highest implantation dose of
$2.00\times10^{17} ions\cdot cm^{-2}$, corresponds to the formation
of the percolated layer of the metal Co nanoparticles in the ZnO
layer.

\begin{figure}[!h]
\includegraphics[width=1\textwidth]{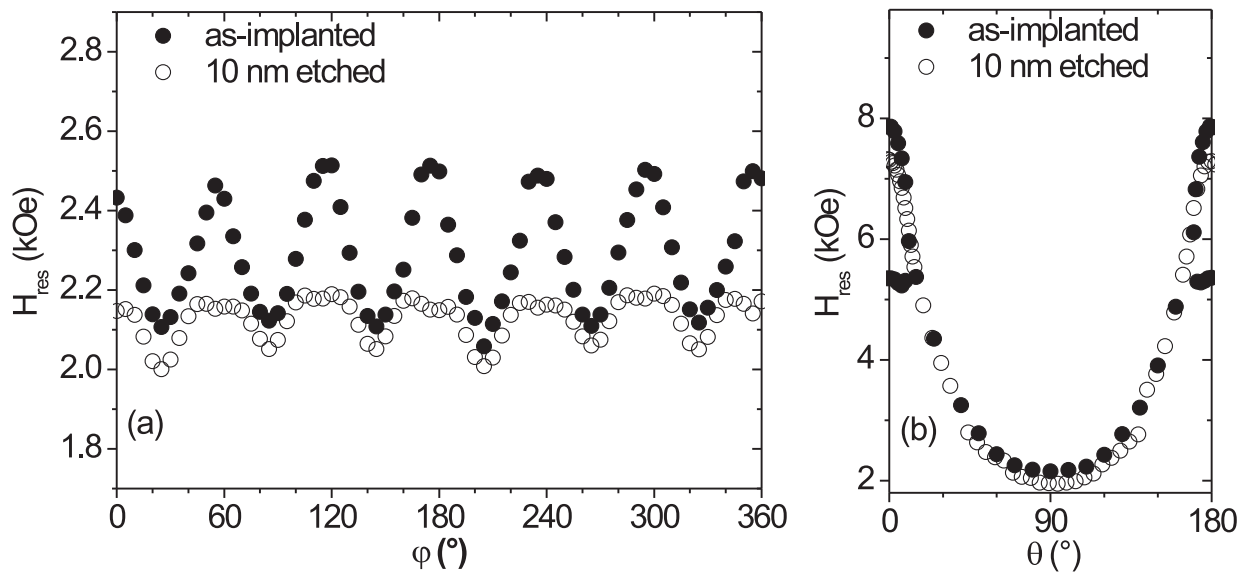}
\caption{\label{4-etching} (a) In-plane magnetic anisotropy of
Co-implanted ZnO film (Co dose: $1.00\times10^{17} ions\cdot
cm^{-2}$) before (closed symbols) and after (open symbols) etching.
(b) Angular dependence of out of plane spectra taken before (closed
symbols) and after (open symbols) etching for the same sample.}
\end{figure}

In order to check whether the contribution to the six-fold in-plane
magnetic anisotropy originates from ZnO only, we gradually removed
the ZnO layer by 500 eV Ar-beam etching and repeated the FMR
measurements. Fig.~\ref{4-etching} presents the in-plane magnetic
anisotropies of Co-implanted ZnO film (Co dose: $1.00\times10^{17}
ions\cdot cm^{-2}$) before and after etching process. When the ZnO
layer is etched by about 10 nm, the six-fold symmetry of the
in-plane magnetic anisotropy survives, but with a significant
decrease in the magnitude of anisotropy. Besides, the signal
intensity also gradually decreases with etching. For the completely
etched sample the signal with six-fold symmetry completely
disappears. This observation shows that the FMR signal with the
six-fold in-plane magnetic anisotropy originates from the
substituted cobalt in the ZnO film, but not from the cobalt
nanoparticles in the Al$_2$O$_3$ substrate.

In conclusion, the magnetic anisotropies of the Co-implanted ZnO
films have been investigated by ferromagnetic resonance technique.
The six-fold in-plane magnetic anisotropy of the FMR signal has been
observed for the first time in ZnO (0001) thin films implanted by Co
in the dose range of $1.00-1.50\times10^{17} ions/cm^{2}$. This
signal is attributed to the ferromagnetic phase formed due to
long-range ordering of substitutional cobalt ions in the ZnO host
matrix. We consider this finding as a strong indication for
intrinsic ferromagnetism in ZnO-based diluted magnetic
semiconductors.
\\

We would like to acknowledge S. Erdt-B\"{o}hm for help in the sample
preparation. This work was partially supported by DPT (State
Planning Organization of Turkey) through the project No 2009K120730,
DFG through SFB 491, by RFBR through the grant No 07-02-00559-a, and
NK-16P/13 of the Federal Agency on Education, Russia.

\end{document}